\documentclass[twocolumn,aps,showpacs,prl,tightenlines,amsmath,amssymb]{revtex4}
\usepackage{graphicx}
\usepackage{amssymb}
\usepackage{amsmath}
\usepackage{colordvi}
\begin{document}
\title{Electron spin rephasing in $n$-type (001) GaAs quantum wells}

\author{K. Shen}
\affiliation{Hefei National Laboratory for Physical Sciences
  at Microscale and Department of Physics, University of Science and Technology of
  China, Hefei, Anhui, 230026, China}
\author{M. W. Wu}
\thanks{Author to whom correspondence should be addressed}
\email{mwwu@ustc.edu.cn.}
\affiliation{Hefei National Laboratory for Physical Sciences
  at Microscale and Department of Physics, University of Science and Technology of
  China, Hefei, Anhui, 230026, China}
\date{\today}
\begin{abstract}
  We investigate the electron spin relaxation in the presence of spin phase
  recovering by a serried $\pi$-pulse sequence in $n$-type (001) GaAs quantum
  wells in a wide range of temperature and density regimes. Our numerical
  calculation is based on the kinetic spin Bloch equation approach with all
  the relevant scatterings explicitly included. We find that the
  rephasing pulse sequence with a long inter-pulse
  spacing only has a marginal influence on the spin lifetime in both the strong
  and weak scattering limits.
  We show that the spin lifetime can be significantly increased by reducing the
  inter-pulse spacing. More
  interestingly, we show that the temperature and density dependences of the
  spin lifetime in the case of short inter-pulse spacing
  coincide with those of the momentum scattering time in the low
  temperature regime, where nonmonotonic behaviors can appear. 
  The origin of this
  feature is that the scattering under the quick rephasing manipulation
  mainly performs as the source of the relaxation channel
  instead of the key to suppress the inhomogeneous broadening.
  The contributions of the relevant scattering mechanisms are also discussed.

\end{abstract}
\pacs{42.50.Hz, 78.67.De, 71.70.Ej}

\maketitle

In the past decade, the electron spin degree of freedom in semiconductors has
been extensively investigated for the
purpose of realizing spintronic devices~\cite{meier,wolf,spintronics,handbook,wu,korn}.
To guarantee the information storage and manipulation in such devices, controllable
spin lifetimes are required. In the localized systems, the interference due
to the inhomogeneous broadening, which corresponds to the distinct spin
  precession of different electrons under the local
  hyperfine nuclear field~\cite{meier} and/or exchange field due to the electron-electron
  interaction~\cite{kavokin01}, can contribute to the decay of the total
spin polarization~\cite{Hahn}. This interference
can be removed via rephasing schemes,
which allows us to significantly extend the ensemble spin
lifetime~\cite{ham,Hahn,abragam,yao,greilich,clark,Bluhm,petta,du}.
For the itinerant electrons in $n$-type semiconductors, the inhomogeneous
broadening is mainly from the momentum-dependent spin-orbit coupling~\cite{wu}
and the electron spin
relaxation is dominated by the D'yakonov-Perel' mechanism~\cite{DP},
where the spin lifetime is limited by the interplay between the inhomogeneous
broadening and scattering. In the strong scattering limit (the average
  scattering time is much shorter than the average period of the spin precession
  due to the effective magnetic field from the spin-orbit coupling),
the inhomogeneous broadening in such systems can be suppressed by the scattering
and the spin relaxation time increases with increasing the
scattering strength~\cite{meier,wu}. This leads to the motional narrowing
relation~\cite{meier} and reveals the way to manipulate the spin
lifetime by tuning the scattering strength in semiconductors. 

Recently, Pershin~\cite{pershin}~introduced a spin rephasing scheme based on a
serried periodic $\pi$-pulse sequence to attenuate the inhomogeneous
broadening in two-dimensional
electron gas. He modeled the isotropic elastic scattering and
found a pronounced increase of the spin lifetime with an inter-pulse spacing
shorter than the momentum scattering 
time. Moreover, he showed that the motional-narrowing relation
of the D'yakonov-Perel' spin relaxation mechanism~\cite{DP} can be violated by
the spin rephasing manipulation. It was suggested that the $\pi$-pulse
  sequence could be realized in the experiment by the optical Stark
  effect~\cite{stark,gupta,sweeney,phelps}.
Although the basic properties under the rephasing sequence have been
discussed in that work, it is still of critical importance to understand the
following problems for the real implementation. First of all,
the influence of the spin rephasing sequence in different temperature and
density regimes is unclear, especially for the high density case at low
temperature, where the motional narrowing relation itself is
unsatisfied. Secondly, since only the elastic scattering was considered in
Pershin's paper, the discussion there is reasonable for the low mobility
system at low temperature, where the inelastic scatterings are irrelevant. However,
in various cases, the inelastic scatterings, e.g., the electron-phonon scattering
at high temperatures and the electron-elctron scattering at moderate
temperatures, can be more important than the elastic
electron-impurity scattering,
especially in high mobility samples. Therefore,
all the relevant scatterings should be included and explicitly calculated
by considering the critical role of the momentum scattering time on the proposed phenomena.

In this letter, we study the spin dynamics under a
serried $\pi$-pulse sequence in $n$-type
(001) GaAs quantum wells (QWs) in a wide range of temperature and density
regimes. Our numerical
simulation is based on the microscopic kinetic spin Bloch
equations~\cite{wu,wu02} with all relevant scatterings, such as, 
the electron-electron, electron-optical-phonon, electron-acoustic-phonon and
electron-impurity scatterings, explicitly included. In high mobility
sample where the electron-impurity scattering is irrelevant, we find
that the rephasing process results in a considerable increase of the spin
lifetime only for the inter-pulse spacing of the
order of several picoseconds. We explicitly study the density and temperature
dependences of the spin lifetime with the inter-pulse spacing taken to be 2~ps,
where the features are found to
be quite different from those in the absence of rephasing sequence.
We show that these 
dependences in both strong and weak scattering limits can be well
understood from the variation of the momentum relaxation time due to the
electron-electron and electron-phonon scatterings. In addition, we also discuss
the effect of the electron-impurity scattering in low mobility samples.

The Hamiltonian of the two-dimensional electron gas in semiconductor QWs
can be written by
\begin{equation}
  H={\bf k}^2/(2m)+H_{\rm soc}({\bf k}),
  \label{eq1}
\end{equation}
with ${\bf k}$ representing the momentum of the electron.
The spin-orbit coupling $H_{\rm soc}$ is introduced to describe the
Dresselhaus~\cite{dressel} and/or Rashba~\cite{rashba} terms between spin-up
and -down states. In (001) symmetric GaAs QW with a small well-width, only the lowest
subband is relevant and the Dresselhaus term is dominant, i.e.,
\begin{eqnarray}
  H_{\rm soc}=2\gamma_D\left[k_x(k_y^2-\langle
  k_z^2\rangle)\sigma_x+k_y(\langle k_z^2\rangle-k_x^2)\sigma_y \right],
  \label{eq6}
\end{eqnarray}
with $\sigma_x$ and $\sigma_y$ being the Pauli matrices. And
$\gamma_D=11.4$~eV\AA$^{-3}$ is the Dresselhaus spin-orbit coefficient~\cite{wu07}.

The temporal evolution of the spin polarization then can be explicitly calculated by
numerically solving the kinetic spin Bloch equations of the density
  matrices~\cite{wu,wu02}
\begin{equation}
  \partial_t\rho_{\bf k}=\partial_t\rho_{\bf k}\big|_{\rm
    coh}+\partial_t\rho_{\bf k}\big|_{\rm scat}.
  \label{ksbe}
\end{equation}
The first and second terms on the right-hand side of the equation represent the coherent
and scattering terms, respectively. The detail expressions of Eq.\,(\ref{ksbe})
can be found in 
Refs.\,\cite{wu07} and \cite{wu03}.

Our calculation starts from the
initial distribution produced by circularly polarized spin-pumping process, i.e.,
the Fermi distribution with a small spin polarization along the growth
direction. We first choose remotely
doped $n$-type GaAs QWs to exclude the impurities. The well-width is taken to be
15~nm, where only the lowest subband is
relevant. Following the previous work by Pershin~\cite{pershin}, we apply a
uniformly spaced $\pi$-pulse sequence with the inter-pulse spacing being $\Delta t$.
We assume that each $\pi$-pulse generated by the optical Stark
effect~\cite{stark,gupta,sweeney,phelps}
rotates the spins of all electrons instantaneously along the growth direction by
$180^\circ$. The numerical error is much smaller than 10\%.

\begin{figure}
\centering
\includegraphics[width=6.2cm]{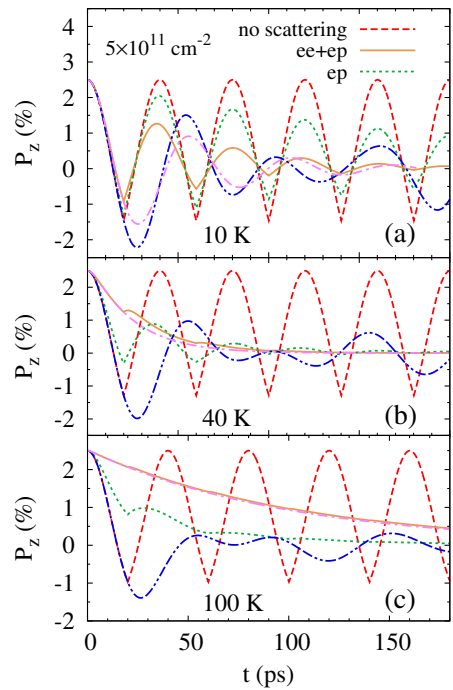}
\caption{ (Color online) Temporal evolution of 
 spin polarization along the growth
  direction in (001) QWs at (a) 10~K,
(b) 40~K and (c) 100~K. The pink chain (blue double-dotted
chain) curves represent the results without rephasing sequence
 in the presence (absence) of scattering. The red dashed and
  khaki solid curves are the results with $\pi$-pulse sequence with $\Delta
  t=40$~ps. The green dotted curves are from the calculation without the
  electron-electron scattering.}
\label{fig1}
\end{figure}

In Fig.\,\ref{fig1}, the temporal evolution of the spin polarization along the
growth direction is plotted
for $N_e=5\times 10^{11}$~cm$^{-2}$ at different temperatures. The blue
double-dotted chain curves show the effect
of the interference without any scattering and rephasing pulse. 
In the presence of
the electron-electron and electron-phonon scatterings, the spin
polarizations vary as shown by the pink chain curves. It is seen that the
oscillation at 10~K is still visible, revealing that the system lies in the
weak scattering limit. However, since the electron gas is in the degenerate
limit ($T\ll T_F$ with the Fermi temperature $T_F\approx 208$~K for $5\times
10^{11}$~cm$^{-3}$) in 
Fig.\,\ref{fig1}, the electron-electron scattering strength as well as the
electron-phonon scattering strength increases with increasing 
temperature~\cite{giulianni}. As a result, the system is pushed into the strong
scattering limit at higher temperatures, e.g., at 40 and 100~K, where the
oscillations are suppressed.

To show the rephasing effect of the $\pi$-pulse sequence, we first take a
large inter-pulse spacing, 
$\Delta t=40$~ps. The results at different temperatures are
shown in Fig.\,\ref{fig1}. In the absence of scattering, the spin polarization
can be completely recovered as shown by the red dashed
curves, because no spin relaxation mechanism exists in this case. When all the relevant
scatterings are switched on, the spin polarization decays due to the 
D'yakonov-Perel'
mechanism (shown as the khaki solid curves). By comparing the results without
rephasing sequence (pink chain curves), one finds that the pulses only have a
marginal influence on the spin dynamics in both the weak and strong scattering
limits. This is because that the
momentum scattering time is 
shorter than the inter-pulse spacing, therefore, the scattering
is the dominant mechanism to suppress the inhomogeneous broadening.
It is noted that the conclusion in the strong scattering limit
 is in agreement with the Monte Carlo
simulation~\cite{pershin}. 
The green dotted curves in Fig.\,\ref{fig1} 
show the results from the calculation solely with
the electron-phonon scattering under the rephasing
manipulation. By comparing with the corresponding ones with both the
electron-electron and electron-phonon scatterings, one finds 
that  the electron-electron
scattering is very important in all the cases~\cite{wuning,glazov02,wu03,jiang}.

\begin{figure}
\centering
\includegraphics[width=4.cm]{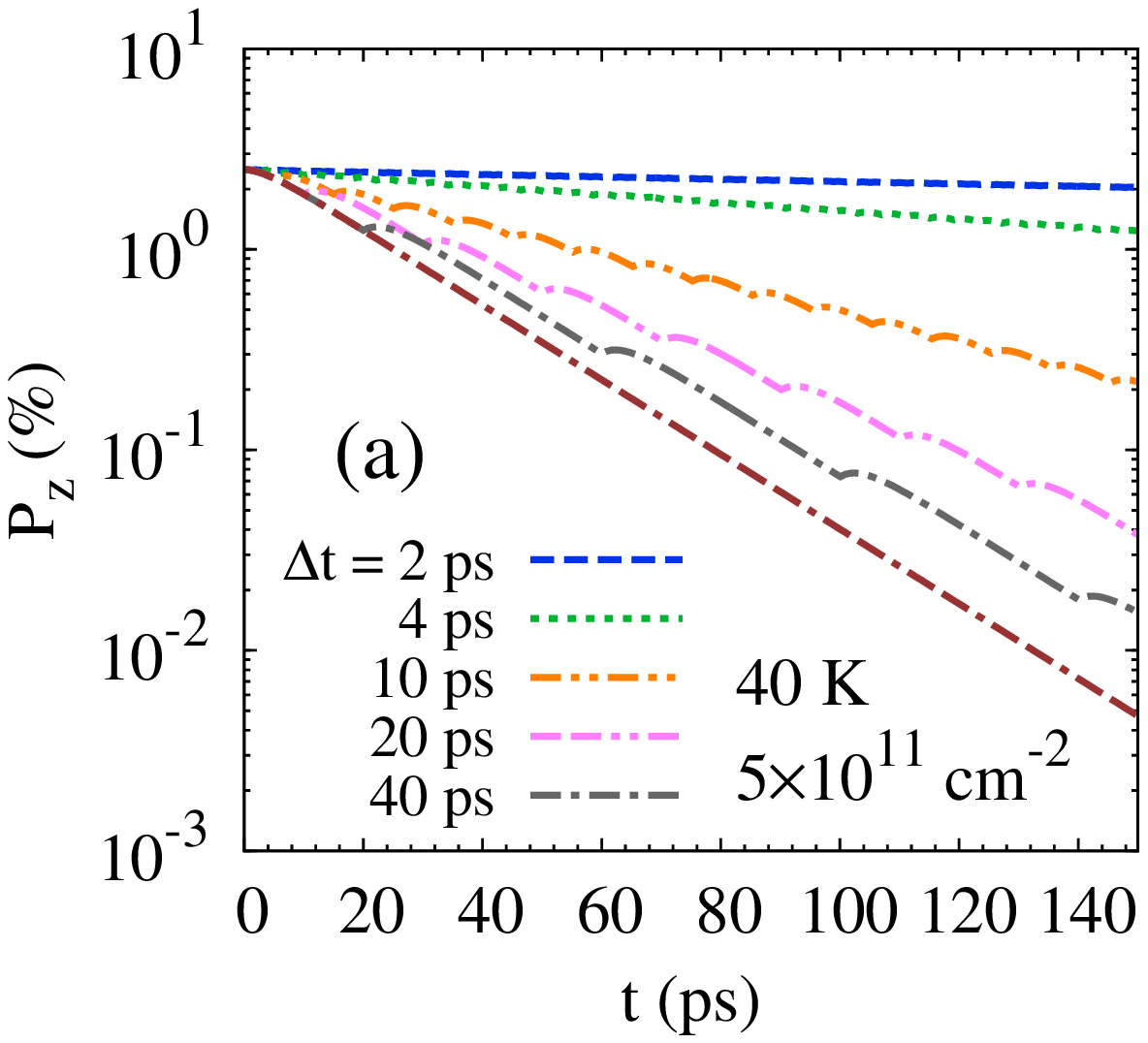}
\includegraphics[width=4.cm]{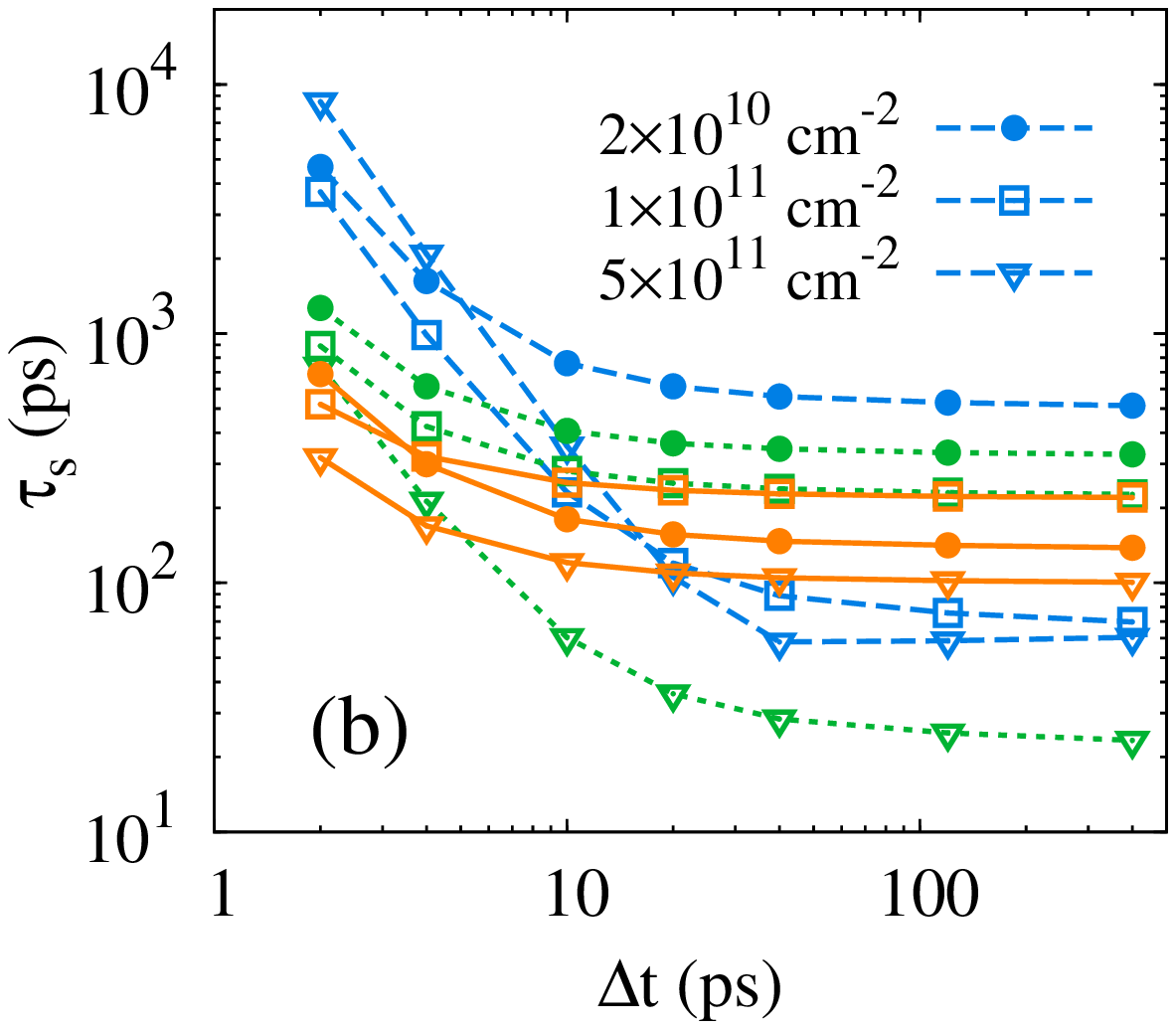}
\includegraphics[width=4.cm]{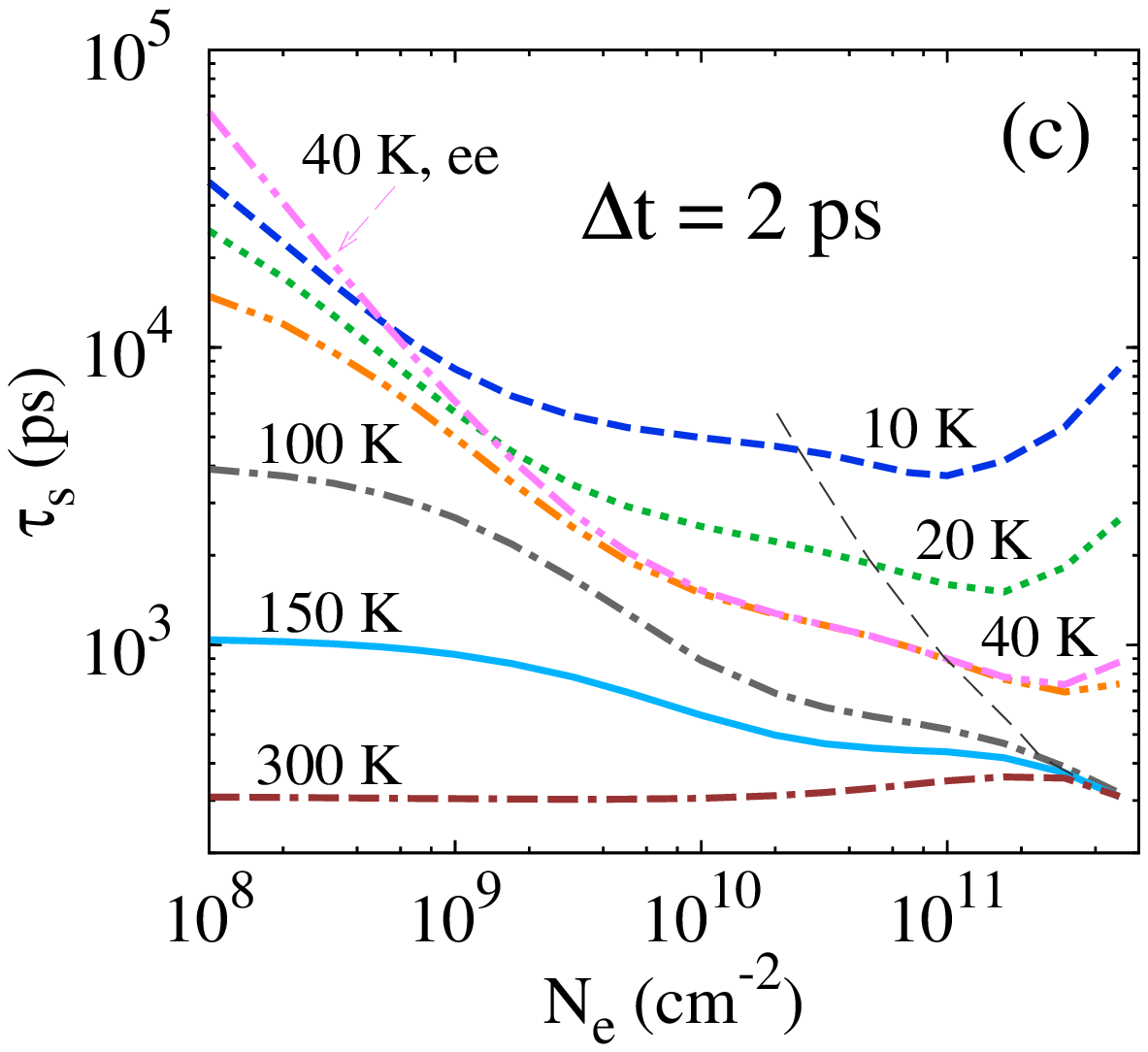}
\includegraphics[width=4.cm]{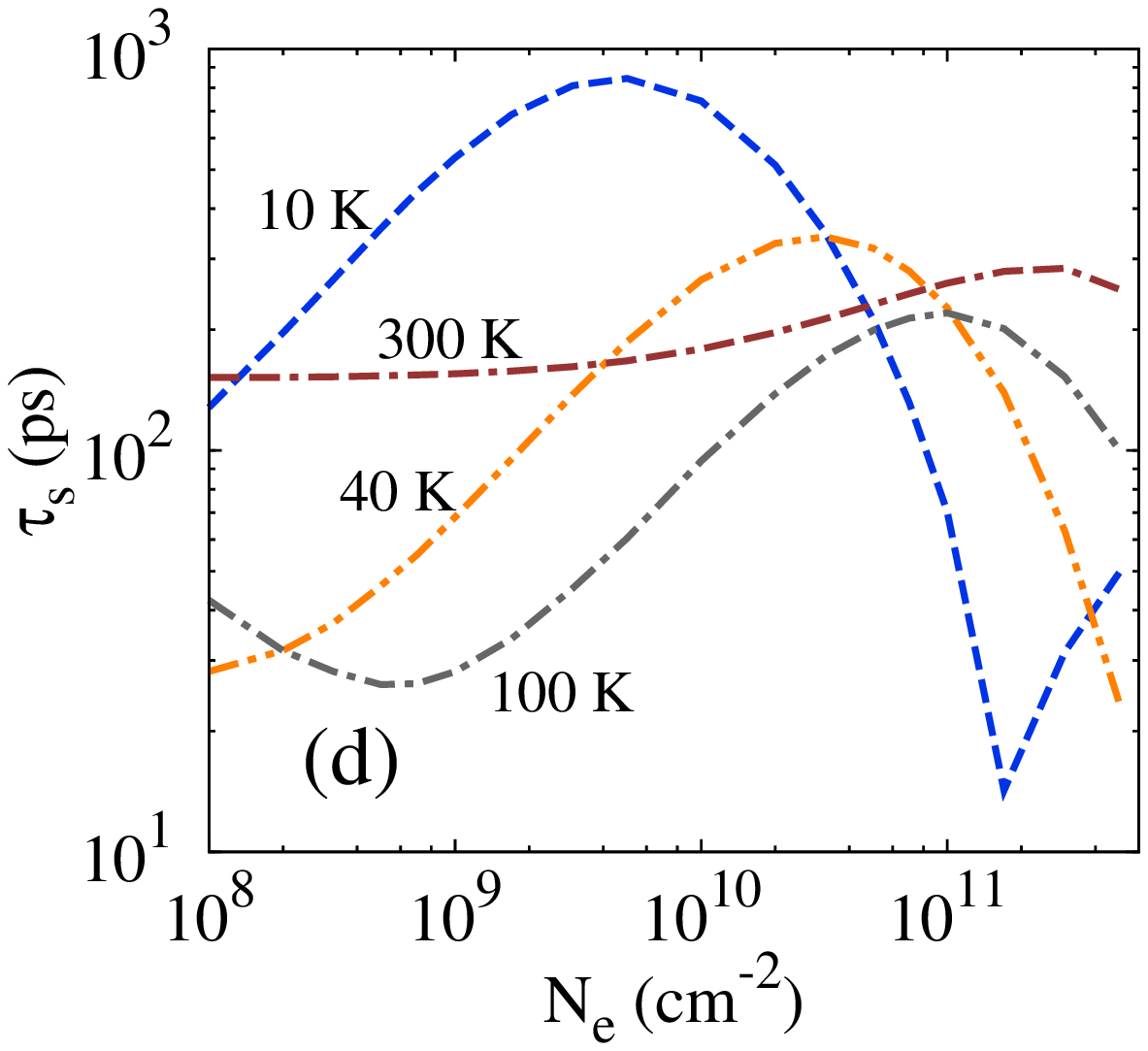}
\caption{(Color online) (a) Temporal evolution of spin polarization 
in the  presence of $\pi$-pulse sequences for
  different inter-pulse spacings. $N_e=5\times 10^{11}$~cm$^{-2}$ at 
40~K. (b) Spin relaxation time as function of inter-pulse
spacing $\Delta t$ at 10~K
  (blue dashed curves), 40~K (green dotted ones) and 100~K (orange solid
  ones). The densities are taken to be $2\times 10^{10}$~cm$^{-2}$ (curves with $\bullet$),
  $10^{11}$~cm$^{-2}$ (ones with $\Box$) and $5\times
  10^{11}$~cm$^{-2}$ (ones with $\nabla$). (c) Spin 
  relaxation time as function of  
  electron density at different temperatures with $\Delta t=2$~ps. The gray
  dashed curve stands for the condition 
  of $T_F=T$. (d) Spin relaxation time as function of density at different
  temperatures in the absence of $\pi$-pulse sequence. The
  curve labeled by ``40~K, ee'' in (c) stands for the result solely due to the
  electron-electron scattering at 40~K, while the others in (a)-(d) are from the
  calculation with both the electron-electron
  and electron-phonon scatterings.
}
\label{fig2}
\end{figure}

We then gradually decrease the inter-pulse spacing and study the inter-pulse
spacing dependence of the spin lifetime in the presence of both the
electron-electron and electron-phonon scatterings. The temporal
evolution of the spin polarization for different inter-pulse spacings at 40~K
is plotted in Fig.\,\ref{fig2}(a). As expected, the spin polarization decays
more slowly
as the inter-pulse spacing decreases~\cite{pershin}. We plot 
the spin lifetime as
function of inter-pulse spacing in Fig.\,\ref{fig2}(b) for three
densities. In this figure, three typical temperatures, i.e.,
10~K (blue dashed curves), 40~K (green dotted ones) and 100~K (orange solid
ones), are chosen. 
In all these cases, the spin lifetime for $\Delta t>40$~ps is found to be
insensitive to the inter-pulse spacing. When $\Delta t$ reduces to 2~ps, the
spin lifetime can even be increased by two orders of magnitude.

We further focus on the short inter-pulse spacing case. In Fig.\,\ref{fig2}(c),
we plot the density dependence under the 
serried $\pi$-pulse sequence with $\Delta t=2$~ps. 
At low temperatures, e.g., at 10~K, one finds that the spin lifetime
decreases (increases) with
increasing density in the low (high) density regime, resulting in a valley in
the density dependence of the spin lifetime. This is in contrast to that at
high temperatures, e.g., at 300~K.
As a comparison, we also show the density dependence of the spin lifetime
in the absence of rephasing sequence in Fig.\,\ref{fig2}(d), where the
peaks are from the crossover between the degenerate and non-degenerate
limits in the strong scattering limit~\cite{jiang2,sun} and the dip
(valley) at $N_e=1.7\times 10^{11}$~cm$^{-2}$ ($N_e=10^{9}$~cm$^{-2}$)
for 10~K (100~K) is due to the crossover between the
strong and weak scattering limits. It is seen that the density dependences of
the spin lifetime with and without 
the rephasing sequence at low temperatures are quite different. 
This is due to the different mechanisms of the suppression of the
inhomogeneous broadening. In the absence of rephasing sequence, the
inhomogeneous broadening is suppressed only by scatterings. However,
when the serried rephasing sequence with inter-pulse spacing being shorter than
the average momentum scattering time ($\Delta t<\tau_p^\ast$) is applied, the
suppression of the inhomogeneous broadening is mainly from the serried
$\pi$-pulse sequence. In such a case, the scatterings only introduce relaxation
channels and one has
$\tau_s\propto\tau_p^\ast$~\cite{pershin}, suggesting that the spin lifetime
varies by following the momentum scattering time.
Below 100~K, the electron-electron scattering is the dominant scattering mechanism.
Therefore, on has
$\tau_p^\ast\sim\tau_p^{ee}\propto N_e/T^{2}$~\cite{glazov,giulianni} in the
degenerate limit, which reveals an
increase of the spin lifetime as the density increases in the high density regime.
However, in the non-degenerate limit, $\tau_p^\ast\propto T/N_e$~\cite{glazov,giulianni} leads
to a decrease of the spin lifetime with increasing density in the low density
regime. This picture is supported by the gray dashed curve in
Fig.\,\ref{fig2}(c) which illustrates the crossover between the
degenerate and non-degenerate limits according to the condition $T_F=T$.
One may notice that the results at 300~K between
Fig.\,\ref{fig2}(c) and (d) show good agreement. This is because that the
enhanced electron-phonon scattering leads to $\tau_p^\ast<\Delta t$. Hence the
scattering becomes the dominant mechanism of the suppression of the
inhomogeneous broadening and the
$\pi$-pulses are irrelevant to the spin relaxation as discussed above.

\begin{figure}
\centering
\includegraphics[width=4.cm]{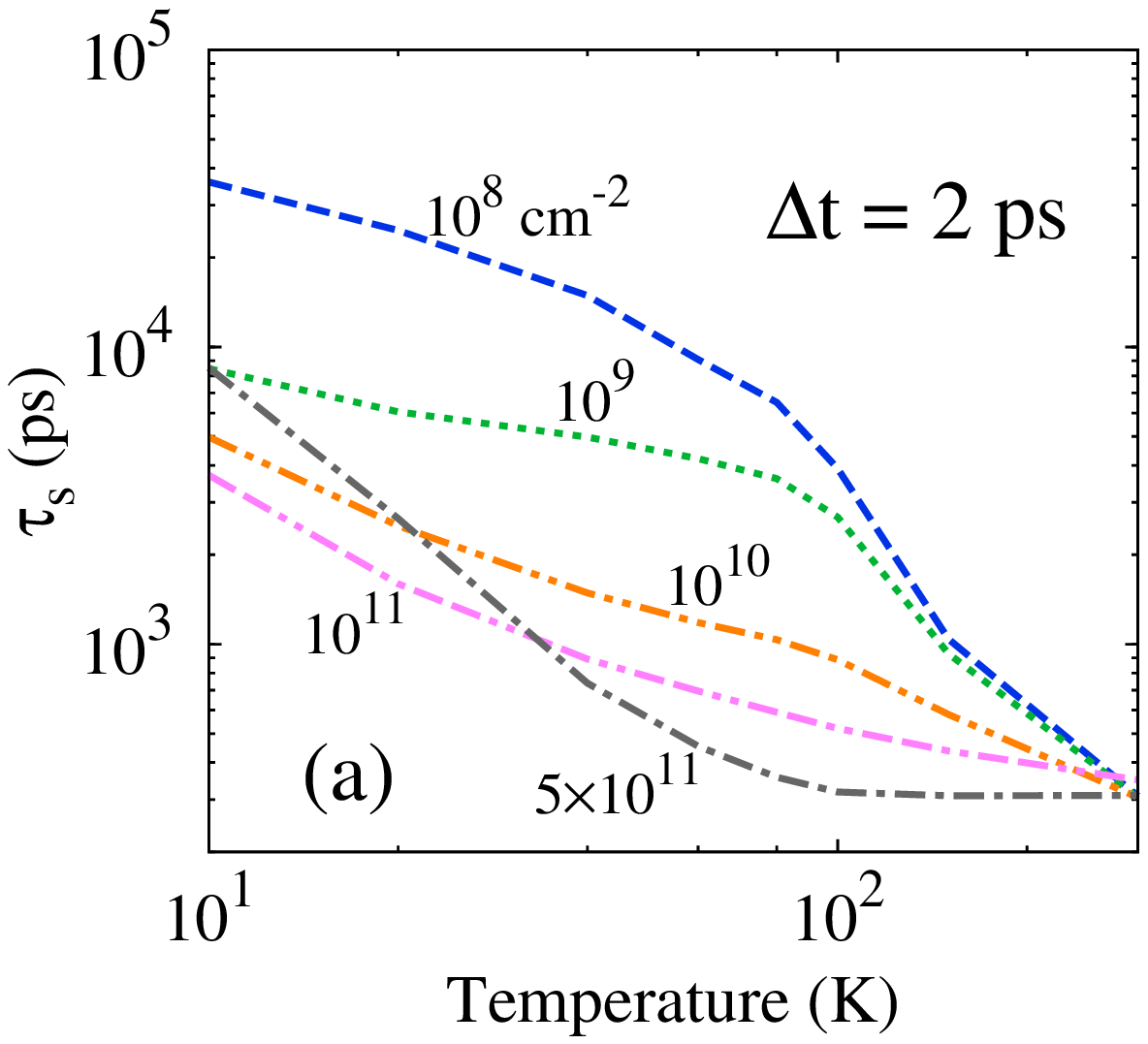}
\includegraphics[width=4.cm]{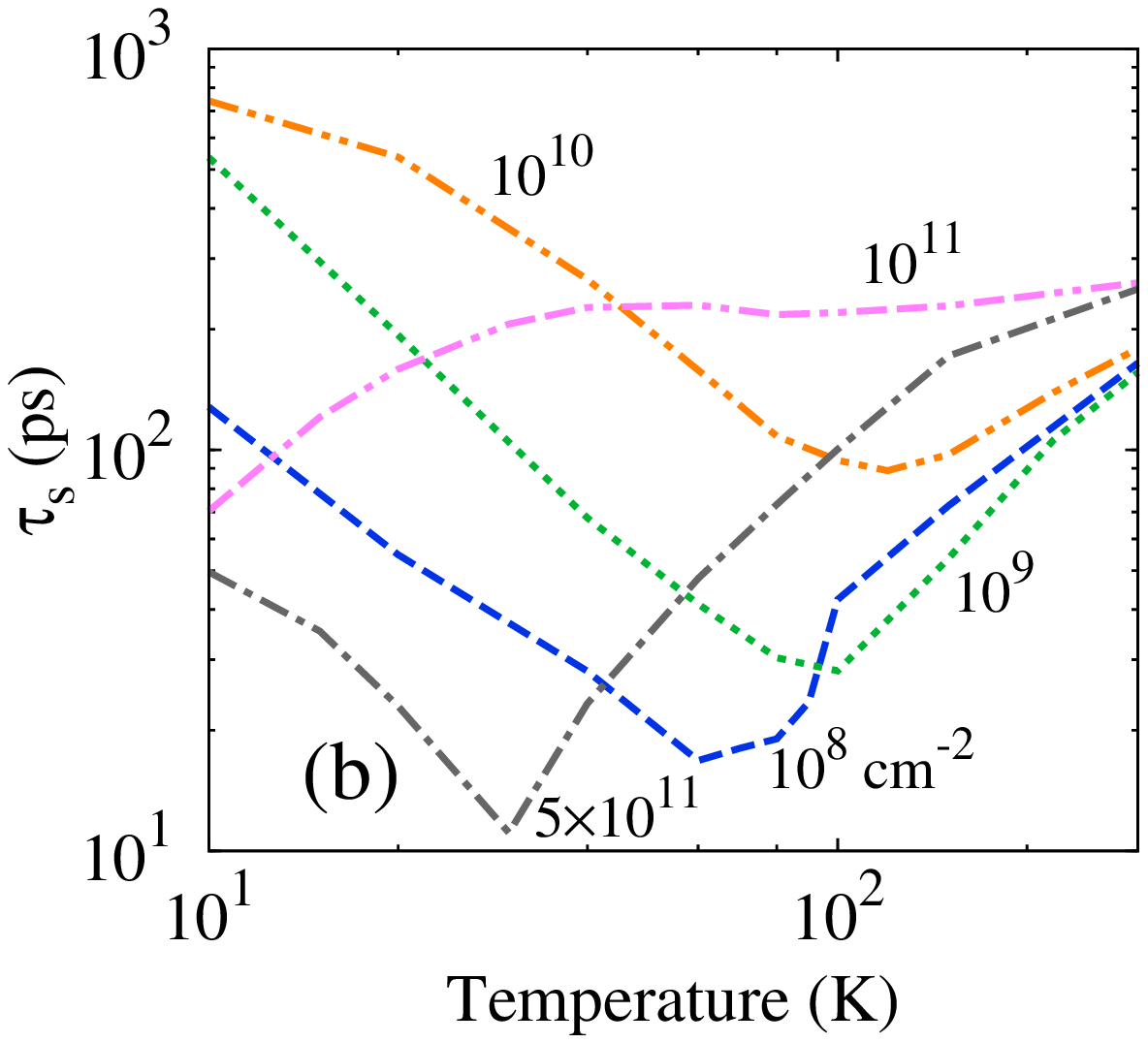}
\caption{ (Color online) Spin lifetime as function of
  temperature (a) with and (b) without $\pi$-pulse sequence at different
  densities in the presence of both the electron-electron and electron-phonon
  scatterings. In (a), the inter-pulse spacing is taken to be $\Delta t=2$~ps.
}
\label{fig3}
\end{figure}

The temperature dependence of the spin lifetime with $\pi$-pulse sequence for
different densities is shown in Fig.\,\ref{fig3}(a), with $\Delta=2$~ps. Since
$\tau_p^\ast>\Delta t$ in the low temperature regime, the spin lifetime
for the high densities, e.g., $5\times10^{11}$~cm$^{-2}$, decreases with
increasing temperature as suggested by the relation 
$\tau_s\propto\tau_p^{ee}\propto N_e/T^{2}$ in the degenerate case.
For the low densities, the system can be in the non-degenerate limit
($T_F\ll T$) even in the low temperature regime. However, one notices
that the spin lifetime still decreases with increasing temperature, violating the relation
$\tau_s\propto\tau_p^{ee}\propto T/N_e$. This is due to the enhancement
of the electron-acoustic-phonon scattering as the temperature increases. In
Fig.\,\ref{fig2}(c), we
plot the spin lifetime from the calculation only with the electron-electron
scattering at 40~K, which is labeled by ``40~K, ee''. One finds
that the spin lifetime in this case can exceed the one at 10~K in
the low density regime (the electron-phonon scattering at 10~K is
negligible in this regime), which supports the relation $\tau_s\propto T/N_e$
solely due to the electron-electron scattering in the non-degenerate limit.
In Fig.\,\ref{fig3}(b), we plot the temperature dependence of the
spin lifetime from the calculation without rephasing sequence to show the
role of the rephasing pulses more clearly. The valleys for
$10^{8}$-$10^{10}$~cm$^{-2}$ are mainly due to the competition between the
inhomogeneous broadening and scattering strength, while the dip at $5\times
10^{11}$~cm$^{-2}$ is
related to the crossover between the strong and weak scattering limits. These
features can be understood by comparing with Fig.\,\ref{fig2}(d).

\begin{figure}
\centering
\includegraphics[width=6.cm]{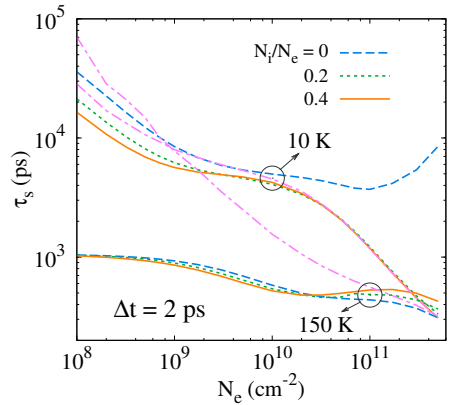}
\caption{ (Color online) Spin lifetime under the influence of the
rephasing pulses  as function of
  density in the presence of impurities at 10 and 150~K. The impurity density
  $N_i$ is taken to be proportional to the electron density $N_e$ with the ratio
  $N_i/N_e=0$ (blue dashed curves), 0.2 (green dotted ones) and 0.4 (orange
  solid ones). The pink chain curves represent the results
with only the electron-impurity scattering with
  $N_i/N_e=0.4$. The inter-pulse spacing $\Delta t=2$~ps.
}
\label{fig4}
\end{figure}

To illuminate the role of the impurity scattering, we introduce
impurities with the impurity densities $N_i/N_e=0$, 0.2 and 0.4. The results are
plotted as function of electron density in Fig.\,\ref{fig4}.
Since one has $\tau^\ast_p>\Delta t$ at 10~K, the increase of the scattering
strength by the electron-impurity scattering reduces the spin lifetime. This is
also the situation in the low density regime at 150~K. 
However, the scattering at 150~K in the high density 
regime is too strong to keep
the relation $\tau_p^\ast>\Delta t$. Therefore, the motional
narrowing relation takes over, resulting in an increase
 of the spin lifetime as the
scattering strength is enhanced by the impurity.
In Fig.\,\ref{fig4}, we also plot the spin lifetime from the calculation
with only the electron-impurity scattering for $N_i/N_e=0.4$ at 10 and 150~K
(shown by the pink chain curves). By comparing
with the corresponding results from the full calculation, one finds that,
  at 10~K, the
electron-impurity scattering is dominant in the high density regime,
while the inelastic scattering is also important in the low density
regime. At 150~K, the enhancement of the electron-phonon scattering makes the
inelastic scattering even dominant.

In summary, we have calculated the spin lifetime due to the
  D'yakonov-Perel' mechanism in two-dimensional electron gas
in (001) GaAs QWs in a wide range of temperature and density regimes
 under a serried
$\pi$-pulse rephasing sequence. We show that the spin lifetime is insensitive
to the rephasing sequence with a long inter-pulse spacing in both the weak and
strong scattering limits. 
As the inter-pulse spacing decreases, the spin lifetime shows a pronounced
increase. In the meantime, the
temperature and density dependences of the spin lifetime can become
quite different from those without $\pi$-pulse sequence due to the different
roles of scatterings. Since the suppression of the inhomogeneous broadening is
mainly from the serried $\pi$-pulse sequence in the short inter-pulse spacing
case, the scattering only performs as the source of the relaxation
channel. We show that
both in the strong and weak scattering limits, the density and temperature
dependences of the spin lifetime under a serried $\pi$-pulse 
sequence are consistent
with those of the momentum scattering time. Moreover, we show that the inelastic
scattering mechanism can present an important contribution even in the presence
of impurities, especially in the low density regime at high temperatures.
According to our results, the 
efficient manipulation of the spin 
lifetime through the $\pi$-pulse rephasing sequence requires high
mobility samples with very low (or high)  electron
densities and low temperatures. Finally, we should point out that the
  extension of the spin relaxation
  time beyond 100~ns can be limited by other spin relaxation mechanisms, e.g., 
  the Elliott-Yafet mechanism~\cite{EY,jiang,jiang2} and/or the fluctuation of the
  hyperfine interaction~\cite{dzhioev}.

\begin{acknowledgments}
This work was supported by the
National Basic Research Program of China under Grant No.\,2012CB922002 and the
National Natural Science Foundation of China under Grant No.\,10725417.
\end{acknowledgments}

\end{document}